\documentclass[sigconf,natbib=true]{acmart}
\usepackage{xcolor}
\usepackage{tablefootnote}
\usepackage{booktabs}
\usepackage{makecell}
\usepackage{float}

\newcommand{\rev}[1]{#1}

\setcopyright{none}
\renewcommand\footnotetextcopyrightpermission[1]{}
\AtBeginDocument{%
  }

\setcopyright{acmlicensed}
\copyrightyear{20256}
\acmYear{2026}
\acmDOI{XXXXXXX.XXXXXXX}
\acmConference[USRW '26]{Unified Search and Recommendation Workshop (USRW), co-located with RecSys '26}{October 2, 2026}{Minneapolis, Minnesota, USA}
\begin{document}

\newcommand{\ml}[0]{\texttt{ML-32M}}
\newcommand{\mli}[0]{\texttt{ML-32M-implicit}}

\newcommand{\mlext}[0]{\texttt{ML-32M-ext}}

\newcommand{\sasha}[1]{\textcolor{red}{#1}}
\newcommand{\newtext}[1]{{\color{purple}#1}}


\title{From IR to RecSys: Evaluating LLM-based Judges in Cranfield-style Recommendation Collections}

\author{Gustavo Penha}
\affiliation{%
  \institution{Spotify}
  \country{United States}
}
\email{gustavop@spotify.com}

\author{Aleksandr V. Petrov}
\affiliation{%
  \institution{Spotify}
  \country{United Kingdom}
}
\email{aleksandrv@spotify.com}

\author{Claudia Hauff}
\affiliation{%
  \institution{Spotify}
  \country{Netherlands}
}
\email{claudiah@spotify.com}

\author{Enrico Palumbo}
\affiliation{%
  \institution{Spotify}
  \country{Italy}
}
\email{enricop@spotify.com}

\author{Ali Vardasbi}
\affiliation{%
  \institution{Spotify}
  \country{Netherlands}
}
\email{aliv@spotify.com}

\author{Edoardo D'Amico}
\affiliation{%
  \institution{Spotify}
  \country{Spain}
}
\email{edoardod@spotify.com}

\author{Francesco Fabbri}
\affiliation{%
  \institution{Spotify}
  \country{Spain}
}
\email{francescof@spotify.com}

\author{Alice Wang}
\affiliation{%
  \institution{Spotify}
  \country{United States}
}
\email{alicew@spotify.com}

\author{Praveen Chandar}
\affiliation{%
  \institution{Spotify}
  \country{United States}
}
\email{praveenr@spotify.com}

\author{Henrik Lindström}
\affiliation{%
  \institution{Spotify}
  \country{Sweden}
}
\email{henok@spotify.com}

\author{Hugues Bouchard}
\affiliation{%
  \institution{Spotify}
  \country{Spain}
}
\email{hb@spotify.com}

\author{Mounia Lalmas}
\affiliation{%
  \institution{Spotify}
  \country{United Kingdom}
}
\email{mounia@acm.org}

\renewcommand{\shortauthors}{Penha et al.}

\begin{abstract}
The Cranfield paradigm has long provided reliable, reproducible evaluation in ad hoc retrieval, and recent work has begun extending this framework to recommender systems. A recent development in IR is the use of Large Language Models (LLMs) as automatic relevance judges, showing promising agreement with human assessors. Whether this LLM-judge paradigm---studied predominantly on query--document pairs---transfers to the subjective, profile-driven nature of recommendation remains an open question. This paper bridges the IR and RecSys evaluation traditions by systematically investigating LLM-based judges within a Cranfield-style recommendation collection. Using the ML-32M-ext movie recommendation collection, we first demonstrate that traditional train--test splits yield substantially incomplete relevance labels and unreliable system rankings compared to Cranfield-style pooling. We then assess LLM-judge alignment with human labels, finding that richer item metadata and longer user histories improve agreement\rev{, although item-level agreement remains moderate overall}. Rankings derived from LLM-judge labels achieve high agreement with human-based rankings (Kendall’s $\tau$ up to 0.92 for nDCG@100 across 52 system configurations), comparable to values reported for TREC ad hoc retrieval collections. Crucially, LLM-judge recovers system rankings that are distorted under traditional evaluation---correctly identifying systems that are undervalued or overvalued by incomplete labels. An industrial case study in podcast recommendation further demonstrates the practical value of LLM-judge for model selection. \rev{Rather than positioning LLM-judges as a replacement for human or interaction-based evaluation, our results support their use as a promising \emph{complementary} signal: item-level agreement with humans is moderate, yet system-level rankings---which aggregate judgments over many user--item pairs---remain stable.}

\end{abstract}

\ccsdesc[500]{Information systems~Recommender systems}
\ccsdesc[500]{Information systems~Relevance assessment}
\ccsdesc[500]{Information systems~Test collections}

\keywords{LLM-judge for recommender systems, Cranfield-style collections, Completeness of relevance labels, Unified search and recommendation evaluation}


\maketitle

\section{Introduction}\label{sec:intro}

Search and recommendation systems share a common goal---connecting users with relevant items---yet their evaluation traditions have developed largely in isolation. In ad hoc retrieval, the \emph{Cranfield paradigm}~\cite{cleverdon1967cranfield} has provided a standardized evaluation framework for decades: test collections of queries, documents, and human relevance judgments enable reproducible and comparable system evaluations, forming the basis of large-scale campaigns such as TREC~\cite{voorhees2019evolution}. Recommender system evaluation, by contrast, still lacks a widely accepted offline framework~\cite{said2014comparative}, relying instead on historical interaction data~\cite{bauer2024exploring}, sampled metrics~\cite{krichene2020sampled,pereira2025reliability}, and heterogeneous splitting strategies~\cite{sun2023take,gusak2025time} that produce inconsistent, often non-reproducible assessments. Issues such as exposure bias~\cite{deffayet2023offline}, popularity bias~\cite{bellogin2017statistical}, and incomplete relevance labels~\cite{rahmani2025judgingjudgescollectionllmgenerated} further complicate the fundamental question: \emph{Which recommendation model performs best}?

Recent work has begun to explore Cranfield-style evaluation frameworks for recommender systems~\cite{chamani2024test,smucker2025extending,umemoto2022ml,hashemi2016overview}, demonstrating that such collections can help mitigate popularity and exposure biases while producing more reliable model rankings. However, these efforts remain costly and limited in scale. For example, extending MovieLens into a Cranfield-style collection required over \$10,000 CAD to obtain relevance labels for only 51 users~\cite{smucker2025extending}, a sample size likely insufficient to capture variance in user preferences and demographics.

In ad hoc retrieval, similar scalability challenges have been addressed by employing  Large Language Models (LLMs) as automatic judges capable of predicting relevance labels. While some concerns remain regarding potential biases, circularity, and robustness to adversarial inputs~\cite{Dietz_2025,10.1145/3673791.3698431}, this approach has often shown high agreement with human assessors~\cite{faggioli2023perspectives,thomas2024large,arabzadeh2025benchmarking,upadhyay2024large,rahmani2025judgingjudgescollectionllmgenerated}. So far, however, this LLM-judge evaluation paradigm has been validated mostly on ad hoc retrieval, where relevance judgments are made for query-document pairs based on the semantic alignment between an information need (expressed through the textual query) and a corresponding document.


In contrast to the relatively objective nature of relevance in retrieval, relevance in recommendation is inherently subjective~\cite{schedl2025psychological}, depending on both context~\cite{mateos2024systematic} and domain~\cite{sun2025we}. Unlike short, well-defined textual queries, recommendation tasks involve long and often noisy user profiles, represented by user-item interactions that reflect evolving preferences over time, which are more challenging to encode for LLMs. Moreover, signals such as popularity, trendiness, and collaborative filtering patterns may be underrepresented in an LLM’s parametric knowledge~\cite{penha2020does}. A related line of work has explored LLMs as user simulators for recommendation (see Section~2), generating synthetic user feedback rather than acting as relevance judges; while complementary, this paradigm differs from the explicit relevance judgment task we study here. These differences raise an important question: \emph{Can the LLM’s judging ability be generalized from ad hoc retrieval to recommendation tasks?}

In this paper, we investigate whether LLM-judge can serve as a reliable evaluator for recommender systems within a Cranfield-style evaluation framework. Our work directly addresses the question of whether evaluation methodologies validated in the IR community can be transferred to RecSys, contributing to the broader goal of unified evaluation across search and recommendation. We begin by motivating our work through an examination of the limitations of existing evaluation methodologies. We then explore: (1) the alignment between LLM-predicted and human-provided relevance labels; (2) the extent to which model rankings based on LLM labels agree with those derived from human judgments; and (3) the performance of LLM-judge in an industrial model-selection setting. We find that:
\begin{enumerate}

    \item[(1)] {\bf Alignment with human labels:} Agreement between LLM-judge and human annotations increases when additional metadata fields (beyond the item title) are provided, suggesting that richer contextual information helps the model better match user preferences to recommended items. We also observe that incorporating more items from a user’s history as input further improves alignment. \rev{Absolute item-level agreement nonetheless remains moderate (55--57\% pairwise agreement) and decreases further for fine-grained interest distinctions.}

    \item[(2)] {\bf Agreement in system rankings:} Using LLM-judge relevance labels to rank different systems yields a Kendall’s $\tau$ agreement of up to 0.92 for nDCG@100 across 52 system configurations, comparable to the alignment scores reported by~\citet{arabzadeh2025benchmarking} for ad hoc retrieval collections such as~TREC-DL. The LLM-judge also correctly recovers system rankings that are distorted under traditional train--test evaluation due to label incompleteness.

    \item[(3)] {\bf Industrial applicability:} In an industrial setting, LLM-judge proves useful for assisting with model ranking and selection prior to conducting human feedback sessions and A/B tests, as demonstrated in podcast recommendation.

\end{enumerate}

Overall, our results indicate that the LLM-judge evaluation paradigm, already validated in IR, \rev{shows promise when applied to recommender systems in domains such as movies and podcasts. We deliberately frame LLM-judges as a \emph{complementary} evaluation signal rather than a replacement for human or interaction-based assessment: item-level agreement with humans is moderate, and LLM-based evaluation carries its own risks of hallucinated reasoning and inherited biases, which we discuss in Section~\ref{sec:discussion}. Established} best-practice guidelines~\cite{Dietz_2025} \rev{should be} followed to ensure validity and integrity of the assessments.

\section{Related Work}

The evaluation paradigms of ad hoc retrieval and recommender systems share deep structural similarities, yet have developed largely independently---a gap that motivates the broader unified search and recommendation research agenda. In ad hoc retrieval, the Cranfield paradigm provides a standardized framework based on test collections composed of documents, queries, and human relevance judgments. Because exhaustively judging every query–document pair is infeasible, large-scale evaluation campaigns such as TREC introduced pooling, where assessors label documents drawn from the union of the top results from multiple retrieval systems.

The completeness of these pooled judgments---particularly for systems that did not participate in pooling---determines the reliability and reusability of a collection for future evaluations. Unpooled documents returned by non-participating systems may still be relevant, yet are treated as irrelevant by standard information retrieval metrics, potentially underestimating the effectiveness of newer models~\cite{arabzadeh2022shallow,voorhees2022too}. Indeed, many commonly used evaluation measures are not robust to substantially incomplete relevance judgements~\cite{buckley2004retrieval}, and the choice of metric itself can significantly affect the reliability of system comparisons~\cite{valcarce2020assessing}. In top-N recommendation, \citet{valcarce2018robustness,valcarce2020assessing} demonstrated that nDCG at deeper cut-offs (e.g., nDCG@100) offers the highest discriminative power and robustness against sparsity, while \citet{otero2025limitations} showed that high system-level rank correlations can mask high rates of false positives regarding statistical differences among top-performing systems.

In recommender systems, the situation is even more complex. Recommendation collections mirror IR test collections---comprising a corpus of items, a set of users, and user–item relevance signals---but these signals are derived from sparse historical interaction data rather than explicit human annotation.\footnote{The relevance of an item to a user can be determined either through pre-interaction judgments, implicit or explicit signals during interaction, or post-interaction feedback~\cite{sun2025we}.} While prior work suggests that assessors can serve as reliable judges on behalf of other users~\cite{lu2021standing,bailey2008relevance}, the standard practice continues to rely on offline evaluations using historical logs. 

Typically, such evaluations withhold a portion of interactions for testing while using the remainder for training (e.g., global-time, leave-one-out splits, or user-based splits)~\cite{10.1007/s10791-020-09371-3,Ji_2023}. However, this methodology is prone to multiple biases: popularity bias arising from long-tail relevance distributions~\cite{bellogin2017statistical}; exposure bias driven by production logging policies~\cite{deffayet2023offline}; and missing-not-at-random (MNAR) patterns~\cite{little2002statistical,steck2010training,schnabel2016recommendationstreatmentsdebiasinglearning}, where users rate only items they particularly enjoy. Additionally, not all rejected recommendations are truly irrelevant~\cite{10.1145/3320435.3320448}. Combined with data leakage from non-time-based splits~\cite{gusak2025time} and the use of sampled metrics~\cite{krichene2020sampled,pereira2025reliability}, these issues can lead to inconsistent evaluations, where offline rankings diverge from those observed in online or user studies~\cite{10.1145/2959100.2959176}, complicating the progress in the field~\cite{ferrari2019we}.

To address these limitations, several efforts have sought to construct Cranfield-style collections for recommender systems, replicating the benefits of IR-style pooling~\cite{smucker2025extending,umemoto2022ml,hashemi2016overview}. Yet, such datasets remain expensive and small in scale. In ad hoc retrieval, LLMs have been shown to effectively fill judgment gaps~\cite{MacAvaney_2023} and predict relevance labels directly~\cite{arabzadeh2025benchmarking}, though concerns persist regarding biases, circularity, and robustness to adversarial manipulation~\cite{Dietz_2025,10.1145/3673791.3698431}. Early work in recommendation further suggests that LLM-judge~\cite{wu2024recsysarenapairwiserecommender,Fabbri_2025} (and LLM rankers~\cite{Dai_2023,hou2024largelanguagemodelszeroshot}) can mirror offline evaluation trends. However, questions remain about their generalization, particularly given evidence that LLMs may memorize public datasets such as MovieLens~\cite{Di_Palma_2025}, and the absence of studies replicating Cranfield-style recommendation collections with LLM judges.

A related line of work explores LLMs as user simulators for recommender systems. Recent surveys~\cite{ni2026survey_user_sim,peng2025survey_agents_recsys} document the growing use of LLM-powered agents that simulate user behavior for training and evaluating recommender systems. \citet{ebrat2025lusifer} propose Lusifer, an LLM-based feedback environment for online recommender evaluation. While user simulation focuses on generating synthetic interactions to train or test recommenders, our work uses LLMs in a different role: as relevance judges that predict human interest labels within a Cranfield-style evaluation framework. More broadly, the intersection of LLMs and recommender systems has been extensively surveyed, covering generative recommendation~\cite{deldjoo2024genrecsys,zhao2024recsys_era_llms} and LLM-based recommendation assistants~\cite{huang2025nextgen_recsys}. Our work is complementary, focusing specifically on the evaluation side rather than the recommendation task itself.

To explore these open questions, the next section outlines our experimental design for evaluating LLM-judge in a Cranfield-style recommender system setting.

\section{Experimental Setup}

In this section, we describe the experimental setup used to evaluate LLM-judge, including the datasets, recommender system models, and evaluation procedures employed.

\paragraph{Pooling-based evaluation for recommendation}
To overcome the limitations of traditional train-test splits, we rely on a Cranfield-style extension of the popular MovieLens interaction dataset~\cite{10.1145/2827872}. The \ml{} dataset\footnote{\url{https://grouplens.org/datasets/movielens/32m/}} consists of 32 million ratings across 87,585 movies by 200,948 users, collected between 1995 and 2023 and released in 2024.
The Cranfield-style extension of the \ml{}, introduced by \citet{smucker2025extending} and referred to as \mlext{}, was released in 2025 and is available exclusively to researchers. Access requires submitting a form confirming that the dataset will not be publicly shared nor used beyond research purposes. Consequently, we consider it unlikely that current LLMs have memorized this dataset.

\mlext{} consists of an additional 31,236 relevance judgments provided by 51 participants for movies that they had not rated before. The movies selected for evaluation were drawn from a pool of 22 recommender system models, each having access to the histories of the 51 participants and other users from the \ml{} dataset. 
During two labeling phases, each participant evaluated at least 600 movies based on metadata including the poster, title, plot summary, release year, running time, cast, directors, genres, and languages. Participants then rated their interest in watching, familiarity with, and predicted rating for each movie.

We use the primary relevance file from \mlext{}: \textit{interest.qrels}, which captures participants’ interest in watching movies. Interest levels range from 0 (Not interested), 1 (Somewhat interested), 2 (Interested), 3 (Very interested), 4 (Extremely interested), 5 (ranked 3), 6 (ranked 2), to 7 (Ranked 1), where 5–7 represent the top-three items selected by each user. For details on dataset construction, we refer readers to~\cite{smucker2025extending, chamani2024test}.

To the best of our knowledge this is the only suitable dataset for investigating LLM-judges Cranfield-style collections in recommender systems. 

\paragraph{Recommender system models}
Following~\citet{smucker2025extending}, we use the RecBole framework~\cite{recbole[1.0]}\footnote{We use RecBole (\url{https://github.com/RUCAIBox/RecBole}) because it provides access to a wide range of recommendation models through a unified API. While we acknowledge existing concerns about the reproducibility of third-party model implementations~\cite{Hidasi_2023_effect}, these have limited impact on our results. Our focus lies in analyzing the alignment between LLM-judge predictions and human judgments, rather than the absolute effectiveness of individual models. Hence, the diversity of models available in the framework is more critical to our study than their specific performance levels.}
for model implementation. Our 52 system configurations span 18 distinct model families: non-personalized baselines (Pop, Random), linear models (EASE~-- Embarrassingly Shallow Autoencoders, ADMMSLIM, SLIMElastic, NCEPLRec), matrix factorization (BPR, NeuMF, ENMF, SimpleX), autoencoders (CDAE, MultiVAE~-- Multinomial VAE, MultiDAE, RaCT, DiffRec), neighbor-based (ItemKNN, AsymKNN), and graph-based (DGCF). Hyperparameter and seed variants yield the full set of 52 configurations. For models that participated in the pooling stage, we adopt the same hyperparameters as in~\cite{smucker2025extending}; for the rest, we train for 10 epochs using default settings. Models with closed-form solutions (e.g., EASE, ADMMSLIM, SLIMElastic) require only a single epoch.
As recommended by the \mlext{} authors, we evaluate system effectiveness using the  Compatibility measure~\cite{DBLP:journals/tois/ClarkeVS21}, with the default value of $p=0.95$ provided by \textit{ir-measures}~\cite{DBLP:conf/ecir/MacAvaneyMO22a} library and the \texttt{interest.qrels} file which contains interest in watching relevance labels. The Compatibility metric is the maximum normalized rank biased overlap between the results and an ideal ranking where $p$ represents searcher patience or persistence.


Full retrieval is performed over the entire item collection, that is no sampled negatives are introduced for metric computation. Items that are not present in the pooling of \mlext{} are considered irrelevant in the evaluation metrics.

For training, we sample 10,000 users from \ml{}, add the history of the 51 \mlext{} participants, and exclude items and users with less than 20 interactions. Preliminary experiments confirmed that this smaller training set yields comparable performance and model rankings on \mlext{}, enabling more efficient experimentation.

\paragraph{LLM-judge}
For the LLM-judge, we use \textit{gpt-5-2025-08-07} with the default reasoning level set to \textit{medium} as the LLM backbone.\footnote{We additionally ran two representative experiments from Sections~5.1 and~5.2 using \texttt{Qwen2.5-32B-Instruct}. While this model cannot handle very long user profiles due to its smaller context length, we observe similar trends: (i) increasing user profile length improves alignment, albeit with 2--4\% fewer agreement pairs, and (ii) using more predictions per user to construct the relevance labels file leads to higher system ranking agreement, with comparable Kendall-$\tau$ scores.} Each experiment is repeated three times, and we report averaged results. Unless stated otherwise, the LLM predicts relevance for up to 50 randomly selected movies per participant (all of which have human labels in \mlext{}). This sampling is adopted to reduce time and costs of experimentation, since a single run would otherwise require 93,708 (3*31,236) LLM calls, and allowed us to run each single point in this experiment three times. The small variance observed across runs with different samples combined with experiments increasing the number of selected items reveal that this does not affect the findings of the paper.

By default, we include 1000 movies for each user as part of their interaction history, along with the user’s explicit rating (on a 1--5 scale) for each movie. Because timestamp information is unavailable for \mlext{}, we randomly sample items from each user’s history rather than applying recency-based selection. The prompt used for LLM-judge evaluations is shown in Figure \ref{fig:llm_prompt}. The prompt was not iteratively optimized, and it is only different for the ablation experiments presented later (see the effect of movie metadata fields in Table~\ref{tab:metadata_ablation}).

\begin{figure}[]
\centering
\fbox{%
\begin{minipage}{0.92\linewidth}
\small
\textbf{LLM-judge prompt}

\vspace{0.5em}
\textbf{Instruction:} \\
Your task is to evaluate the relevance of a movie recommendation for a user,
given the metadata of the movie and the user’s previous interactions. \\[0.5em]

\textbf{Inputs:} \\
\textit{User profile:} \texttt{\{USER\_PROFILE\}} \\
\textit{Movie recommendation:} \texttt{\{METADATA\_RECOMMENDED\_MOVIE\}} \\[0.5em]

\textbf{Expected output:} \\
\texttt{reasoning:} Reason for the interest in watching the movie considering their previous interactions. \\
\texttt{interest\_in\_watching:} The interest in watching the movie considering their previous interactions, on a scale of 0 to 7.
\end{minipage}}
\\[1em]
\fbox{%
\begin{minipage}{0.92\linewidth}
\small
\textbf{User profile structure}

\vspace{0.5em}
\texttt{Movie metadata: \{METADATA\_MOVIE\_0\}} \\
\texttt{User rating on scale 1-5: \{RATING\_0\}} \\[0.3em]
\texttt{Movie metadata: \{METADATA\_MOVIE\_1\}} \\
\texttt{User rating on scale 1-5: \{RATING\_1\}} \\[0.3em]
\texttt{\dots} \\[0.3em]
\texttt{Movie metadata: \{METADATA\_MOVIE\_N\}} \\
\texttt{User rating on scale 1-5: \{RATING\_N\}}
\end{minipage}}
\\[1em]
\fbox{%
\begin{minipage}{0.92\linewidth}
\small
\textbf{Movie metadata structure}

\vspace{0.5em}
\texttt{Title + Average rating + Genres + Directors + Overview + Cast + Runtime + Number ratings + Year + Languages.}
\end{minipage}}
\caption{Prompt template, user profile structure, and movie metadata structure used for LLM-judge evaluations.}
\label{fig:llm_prompt}
\end{figure}

\paragraph{\rev{Reproducibility}}
\rev{While we are unable to release code or prediction files, all components required to reproduce our setup are documented in this paper: the dataset and its statistics as well as the recommender system models (this section), the exact prompt template (Figure~\ref{fig:llm_prompt}) and its metadata-ablation variants (Table~\ref{tab:metadata_ablation}), the LLM backbone, settings, and sampling procedure (above), and the hyperparameter and seed variants of all 52 system configurations (Appendix~\ref{app:hyperparams}). Access to \mlext{} must be requested from the original authors~\cite{smucker2025extending}.}

\section{Limitations of Current Evaluation Methodologies}\label{sec:limitations}

Evaluating recommender systems using traditional offline methodologies presents several fundamental limitations. In this section, we examine the incompleteness of relevance labels under different evaluation methodologies and the implications for the reliability of model comparisons.

A key limitation of traditional train-test splits based on historical interactions is the incompleteness of relevance labels: not all items retrieved by the compared recommender systems have corresponding relevance labels. The resulting sparsity, across different models, makes it difficult to reliably compare recommender system rankings, as confirmed by prior studies on partially complete collections~\cite{Gao_2022,pereira2025reliability}. We use Judged@100\footnote{\url{https://ir-measur.es/en/latest/measures.html\#judged}}, defined as the average percentage of items with relevance judgments (any relevance level) among the top 100 ranked items for each user, to quantify this incompleteness.


We next compare the completeness of traditional train–test splits with that of Cranfield-style collections and show how low Judged@100 values lead to discrepancies in system rankings.
\begin{table}[tb]
\centering
\caption{Average Judged@100 under historical interactions train--test split vs. Cranfield-style pooling (\mlext{}).}
\renewcommand{\arraystretch}{1.15}
\setlength{\tabcolsep}{6pt}
\begin{tabular}{@{}lcc@{}}
\toprule
\textbf{Models} & \makecell[c]{\textbf{\ml{}}\\[-2pt]\small Historical 80\%--20\% \\ \small train--test split}
& \makecell[c]{\textbf{\mlext{}}\\[-2pt]\small Cranfield-style \\ \small pooling} \\
\midrule
Pooled models (7)      & 0.0774 & \textbf{0.5981} \\
Non-pooled models (45)  & 0.0601 & \textbf{0.5503} \\
\bottomrule
\end{tabular}
\label{tab:judged100}
\end{table}

\paragraph{Comparison with a Cranfield-style pooled collection}

A Cranfield-style collection offers a compelling alternative to traditional historical train–test splits. By collecting explicit feedback from multiple recommender systems, such collections provide a more complete set of relevance judgments for models that participated in the pooling process, that is, models used to generate candidate recommendations during dataset creation. Incorporating a diverse set of models into the pool also helps mitigate the selection bias that would arise from relying on a single production recommender.

In standard offline evaluation, user interactions are typically sorted by time and then split into train–test sets on a per-user basis~\cite{sun2020we}.\footnote{For example, the widely used RecBole library does not implement a global time-based split and only supports time-ordering prior to splitting the data into training and test sets (see \url{https://recbole.io/docs/user_guide/config/evaluation_settings.html}).} This approach can inadvertently cause temporal leakage, as each user has their own time cutoff between training and testing~\cite{Ji_2023,Hidasi_2023,gusak2025time}. In contrast, Cranfield-style collections enable a natural global time split, where all interactions occurring before the dataset creation are used for training, and the relevance judgments collected during dataset construction are used for testing.



We argue that this setup more closely reflects real-world recommendation scenarios, offering both higher completeness of relevance labels and more reliable model rankings. To quantify this difference, we compare the completeness of relevance labels under the two methodologies.

%
%

As shown in Table~\ref{tab:judged100}, traditional train–test splits yield substantially lower Judged@100 scores compared to the Cranfield-style collection (\mlext{}~\cite{smucker2025extending}). We evaluate 52 system configurations comprising 18 distinct models with hyperparameter and seed variants. Models that participated in the \mlext{} pooling process see a large increase in completeness under the Cranfield-style setup (from 0.08 to 0.60 on average). A similar trend is observed for the 45 system configurations that did not participate in pooling: their Judged@100 values are also higher under the Cranfield-style evaluation (from 0.06 to 0.55), though overall scores are lower than for the pooled models. Notably, the Pop (popularity) baseline achieves low Judged@100 even under Cranfield-style pooling, because its recommendations are drawn from globally popular items that may not overlap with the items pooled from the 22 personalized models used during dataset construction.




\paragraph{Impact on model ranking reliability}

\begin{figure}[t]
    \centering
    \includegraphics[width=1\linewidth]{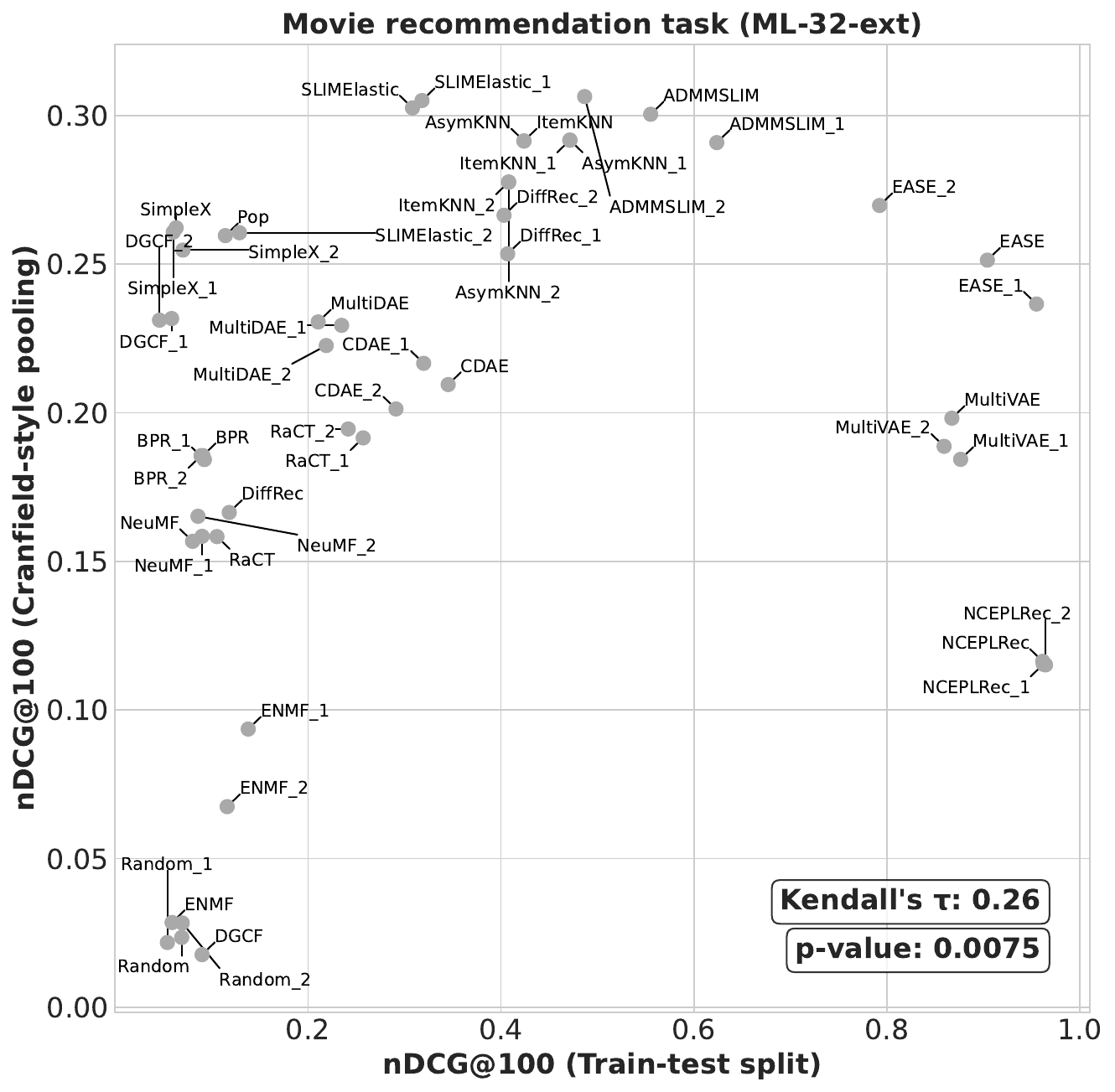}
    \caption{nDCG@100 under the traditional train-test split vs.\ the Cranfield-style pooled collection for 52 system configurations. The low correlation (Kendall’s $\tau = 0.26$) shows that system rankings change substantially with more complete relevance labels.}
    \label{fig:tt_vs_cranfield}
\end{figure}

An important consideration is how the evaluation methodology and the completeness of relevance labels affect the final rankings of recommender systems. Across our 52 system configurations, the agreement between nDCG@100 rankings derived from the historical interactions train-test split (which have low completeness) and those from the Cranfield-style collection (which have higher completeness) is notably low: 0.26 Kendall’s $\tau$ ($p < 0.01$), as shown in Figure~\ref{fig:tt_vs_cranfield}. For comparison, an often cited threshold for acceptable agreement between system rankings based on independently created set of human relevance judgments is 0.90~\cite{voorhees1998variations, arabzadeh2025benchmarking}.

The rank disagreements are not merely statistical noise---they reflect systematic distortions. For example, NCEPLRec is ranked 1\textsuperscript{st} out of 52 configurations under the train-test split but drops to 43\textsuperscript{rd} under the Cranfield collection, a shift of 42 positions. Conversely, SimpleX rises from 47\textsuperscript{th} under the train-test split to 14\textsuperscript{th} under Cranfield (+33 positions), and DGCF configurations improve by up to 29 positions. These ``undervalued’’ runs recommend items outside the user’s historical interactions that human assessors nonetheless rate as relevant, while ``overvalued’’ runs like NCEPLRec and MultiVAE (which shifts by up to 30 positions) primarily succeed at predicting held-out interactions without genuinely capturing user interest.

\paragraph{Limitations of pooled collections}

While pooled collections improve the completeness of relevance judgements, they are inherently``\textit{frozen in time}'': evaluation can only be conducted using a single time-based split (before and after the pooling process). In domains where user preferences evolve rapidly (e.g. short video consumption) reliable evaluation may require multiple temporal splits to capture such dynamics.

In addition, decisions regarding which models participate in pooling, how they are trained, and the pooling depth can influence both the completeness and longevity of the collection. Models that resemble those used during pooling may be inadvertently favored, while newer or substantially different models may be disadvantaged. 
Finally, constructing pooled collections is resource-intensive. For instance, generating \mlext{} with participants required approximately \$10,000 CAD. In contrast, using an LLM-judge could reduce this cost by an order of magnitude.\footnote{The estimate was obtained using \textit{GPT-5 mini}, via the batch API, based on the average context length of a textified user history containing all metadata and 100 movies. Optimizations on the number of LLM relevance predictions and which metadata to use could make the LLM-judge 2--3 orders of magnitude cheaper.}


\paragraph{Summary} 

\begin{table}[ht!]
\centering
\caption{Comparison of evaluation methodologies for recommender systems.}
\begin{tabular}{p{0.30\linewidth} p{0.30\linewidth} p{0.30\linewidth}}
\toprule
\textbf{Historical interactions} & \textbf{Cranfield-style Pooling} & \textbf{LLM-judge} \\
\midrule
Selection bias towards production recommender system(s) and towards popular items &
Selection bias towards systems used during pooling &
No selection/exposure bias;\newline other biases remain.\\ \midrule
\addlinespace
Incomplete labels &
Incomplete labels &
``\textit{Complete}''\tablefootnote{By a complete set of relevance labels, we refer to the ability to run LLM-judge for any item retrieved by any recommender system within the top-k positions, rather than generating relevance predictions for every possible user–item pair in the collection.} labels \\ \midrule
\addlinespace
Low cost &
High cost &
Medium cost \\ \midrule
Low reliability for model ranking &
High reliability for model ranking\tablefootnote{Cranfield-style collections provide high reliability when ranking recommender systems that participated in the pooling process, but their reliability decreases for models that did not participate and differ substantially from those included in the pool.} &
Moderate reliability for model ranking 
\\
\bottomrule
\end{tabular}
\label{tab:evaluation_comparison}
\end{table}

As summarized in Table \ref{tab:evaluation_comparison}, existing evaluation methodologies for recommender systems involve a trade-off between bias, completeness, and cost. Historical-interaction evaluations are inexpensive but suffer from biases and incompleteness, whereas Cranfield-style pooling offers higher reliability at the expense of substantial human-annotation cost. 

The LLM-judge approach provides an appealing middle ground: assuming sufficient alignment with user judgments, it mitigates exposure and  popularity biases by offering complete coverage of relevance labels at a moderate cost. 

However, LLM-based evaluations introduce their own sources of bias~\cite{ye2024justice,sguerra2025biases} and potential pitfalls that may lead to invalid conclusions~\cite{Dietz_2025,clarke2024llm}. For example, popular entities or items may be overrepresented in an LLM's parametric knowledge~\cite{ni2025knowledge, lichtenberg2024large}\rev{, so judgments for long-tail items are likely to be less reliable than for mainstream ones}; test sets might inadvertently appear in the model’s training data~\cite{Di_Palma_2025, xu2024benchmark}; \rev{LLMs may hallucinate---generating confident but factually incorrect reasoning about item attributes or user preferences that results in plausible-looking yet wrong labels;} LLMs can be adversarially prompted to produce overly positive relevance labels~\cite{10.1145/3673791.3698431}; and using LLM-judge within the evaluated systems themselves risks circularity issues~\cite{li2025preference}.

These potential advantages and shortcomings motivate our investigation into whether LLMs can serve as scalable and reliable judges for recommender system evaluation.

\section{Assessing LLM-judge Alignment and Agreement}


In this section, we evaluate the effectiveness of LLM-judge as an automatic evaluator for recommender systems. Following the evaluation methodology of ~\citet{arabzadeh2025benchmarking}, we assess LLM-based relevance judgments along two dimensions: (1) their alignment with human labels, and (2) their agreement with recommender system rankings derived from human relevance judgments. Together, these analyses allow us to understand not only how closely LLM-judge mirrors human annotation behavior but also whether it preserves the relative ordering of model performance, both essential requirements for scalable, reliable evaluation.

\subsection{Alignment with Human Labels} 
We first evaluate the alignment between LLM-judge predictions and the primary relevance file from \mlext{}. The LLM-judge receives a fixed number of  randomly ordered historical user ratings as inputs,\footnote{The LLM-judge has access to the same metadata provided to participants during the relevance labeling process, with the exception of the cover image and the addition of two extra fields: average rating and number of ratings.} enabling it to infer how well a user's preferences align with the metadata of a recommended movie. Both the LLM-judge predictions and the human labels are expressed on the same interest scale ranging from 0 to 7.

To quantify this alignment, we calculate the proportions of agreement, tie and disagreement between label pairs. Specifically, for each user, we compare all pairs of items labeled as relevant (interest levels 1–7) versus non-relevant (interest level 0), and determine:
\begin{itemize}
    \item {\bf Agreement:} the LLM-judge ranks the items in the same order as the human judgments.
    \item {\bf Tie:} the LLM-judge assigns equal scores to both items. 
    \item {\bf Disagreement:} the LLM-judge ranks the pair in the opposite order from the human judgments.
\end{itemize}

\paragraph{Effect of including different item metadata fields}

We first examine how different metadata fields (used for movies in both the user's history and the recommended movie) contribute to the alignment between LLM-judge and human labels.

As shown in Table~\ref{tab:metadata_ablation}, all metadata fields except average rating positively impact the pairwise agreement between LLM-judge predictions and human judgments. Overall, these results indicate that the model effectively leverages the content information from a user’s movie profile and compares it against the metadata of recommended items; each additional metadata field independently helps improve alignment. At the same time, interaction-derived features such as average rating and number of ratings may inadvertently reinforce---or reintroduce---popularity biases into the evaluation.

\begin{table*}[t]
\centering
\caption{Effect of movie metadata fields on LLM-judge pair agreement metrics. ($\pm$) indicates the 95\% confidence interval. Best per column in bold. $\downarrow$ indicates worse than the baseline which only uses the \textit{Title} of the movies as input.}
\renewcommand{\arraystretch}{1.15}
\setlength{\tabcolsep}{10pt}
\begin{tabular}{llll}
\toprule
\textbf{Metadata} & \textbf{Agreement (\%)} & \textbf{Tie (\%)} & \textbf{Disagreement (\%)} \\
\midrule
Title + Average rating & $0.5398 \pm 0.0033^{\downarrow}$ & $0.2565 \pm 0.0008$ & $0.2037 \pm 0.0028^{\downarrow}$ \\
Title                  & $0.5479 \pm 0.0019$ & $0.2607 \pm 0.0017$ & $0.1914 \pm 0.0011$ \\
Title + Genres         & $0.5516 \pm 0.0020$ & $0.2596 \pm 0.0000$ & $0.1888 \pm 0.0020$ \\
Title + Directors      & $0.5551 \pm 0.0006$ & $0.2514 \pm 0.0024$ & $0.1935 \pm 0.0024^{\downarrow}$ \\
Title + Overview       & $0.5529 \pm 0.0019$ & $0.2645 \pm 0.0007$ & $0.1826 \pm 0.0025$ \\
Title + Cast           & $0.5649 \pm 0.0033$ & $0.2456 \pm 0.0008$ & $0.1895 \pm 0.0030$ \\
Title + Runtime        & $0.5592 \pm 0.0005$ & $0.2561 \pm 0.0012$ & $0.1847 \pm 0.0015$ \\
Title + Number ratings     & $0.5612 \pm 0.0022$ & $0.2489 \pm 0.0005$ & $0.1899 \pm 0.0017$ \\
Title + Year           & $0.5590 \pm 0.0011$ & $0.2536 \pm 0.0012$ & $0.1874 \pm 0.0021$ \\
Title + Languages      & $\mathbf{0.5726} \pm \mathbf{0.0008}$ & ${0.2362} \pm {0.0021}$ & $0.1912 \pm 0.0013$ \\
Title + All metadata   & $0.5599 \pm 0.0031$ & $0.2610 \pm 0.0013$ & $\mathbf{0.1790} \pm \mathbf{0.0021}$ \\
\bottomrule
\end{tabular}
\label{tab:metadata_ablation}
\end{table*}






When compared with prior work on LLM-based judges for ad hoc retrieval ~\cite{arabzadeh2025benchmarking}, our alignment scores (55–57\%) are higher than those achieved by pointwise LLM-judges on the ANTIQUE dataset (\textasciitilde40\%) but lower than those observed on TREC DL-19/20/21 (\textasciitilde80\%). It is worth noting that our movie recommendation judge operates in a zero-shot setting without prompt iteration or in-context learning examples, which makes these results particularly encouraging.

\paragraph{Effect of the size of the user history}

When using all available metadata fields for each movie, the context size of the user profile grows rapidly. On average, each user who participated in the \mlext{} pooling rated 1,425 movies, resulting in an average profile length of approximately 170k tokens, with a maximum of 1 million tokens for a user who rated 8,496 movies.
\begin{figure}[t]
    \centering
    \includegraphics[width=1\linewidth]{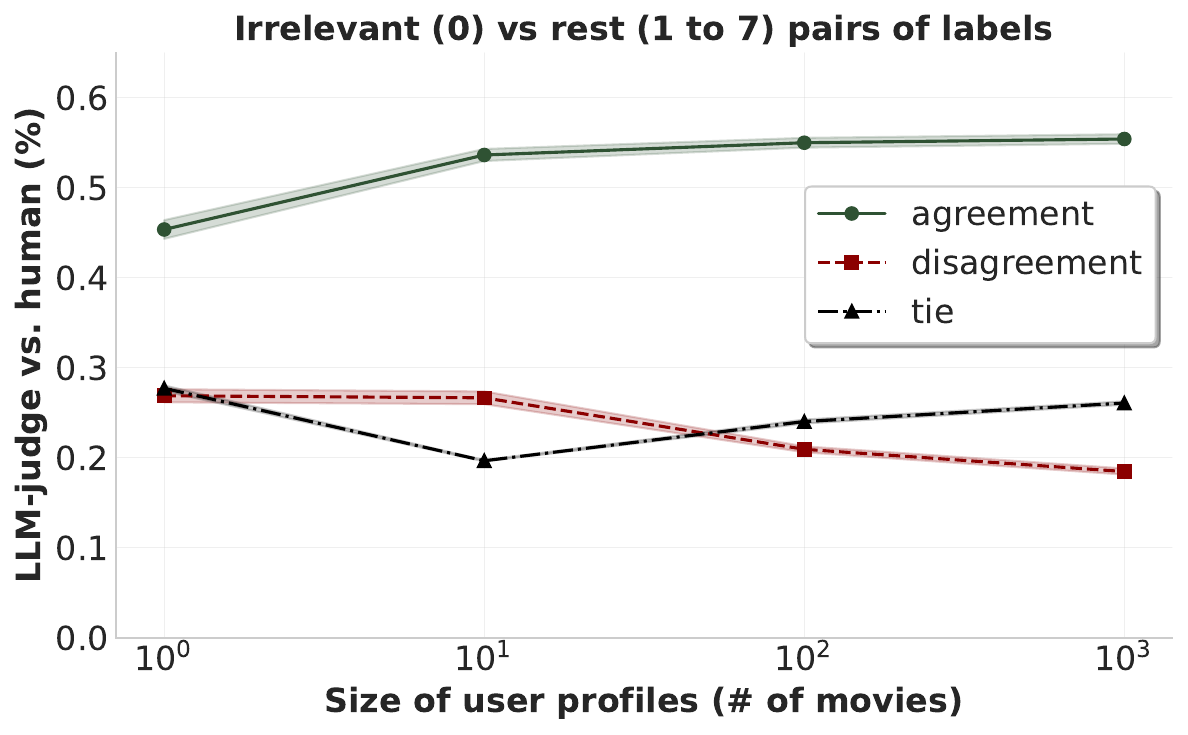}
    \caption{Effect of increasing the size of the user history on the LLM-judge pair agreement.}
    \label{fig:user_history_size}
\end{figure}
To understand the impact of profile size, we analyze how the number of items in a user’s history influences the alignment between LLM-judge and human relevance labels. As illustrated in Figure~\ref{fig:user_history_size}, larger user profiles lead to higher pairwise agreements and fewer disagreements between the two sets of judgments. This trend is consistent with early findings from LLM-judge evaluations in podcast recommendations~\cite{Fabbri_2025}. We note that timestamp information for the user interactions was not available for \mlext{}, and therefore, profile items were not ordered by recency in this analysis. These results suggest that providing richer user context enables LLMs to better infer preference patterns, improving their ability to approximate human relevance judgments.


\paragraph{Effect of the difference in human label interest levels}
Previous sections evaluate the LLM-judge agreement using not relevant (interest level of 0) vs relevant (interest levels from 1--7) pairs of items. Here, we take a closer look at agreement over unrestricted pairs of interest levels, considering more fine-grained distinctions (e.g. interested vs very interested).

\begin{figure}[t]
    \centering
    \includegraphics[width=1\linewidth]{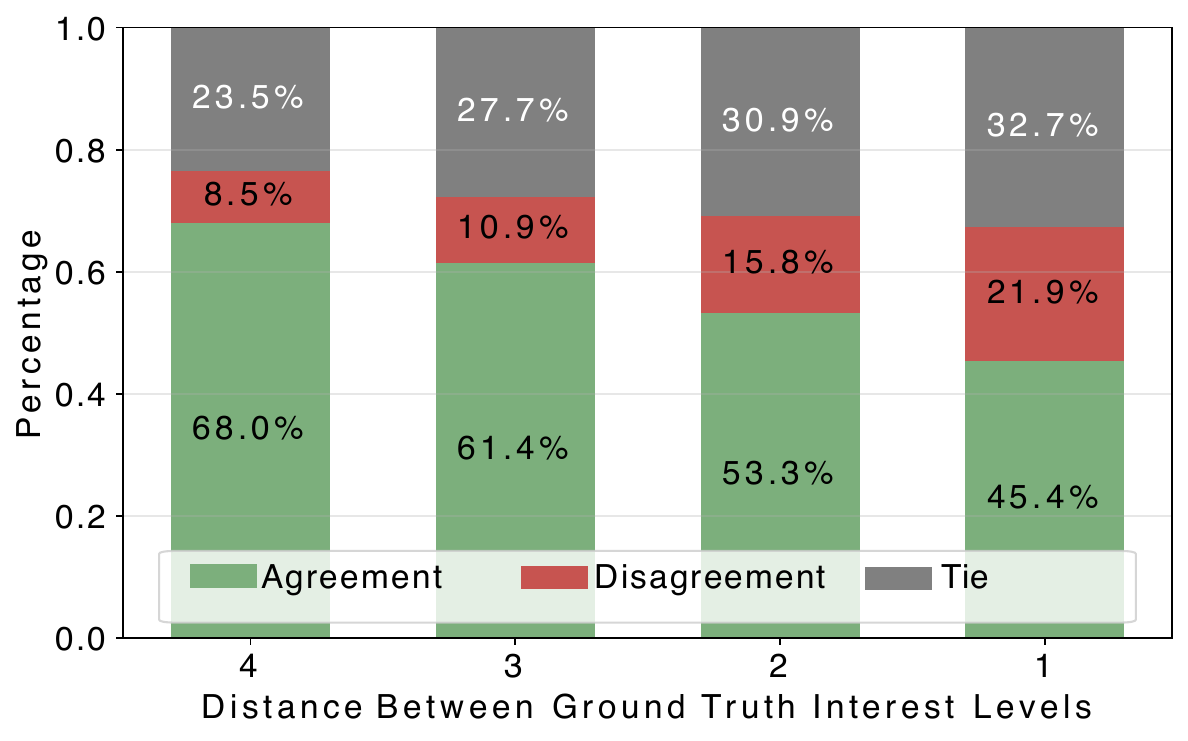}
    \caption{Effect of decreasing the interest level differences between the pairs of labels.}
    \label{fig:agreement_any_pair}
\end{figure}

As shown in Figure~\ref{fig:agreement_any_pair}, agreement decreases as the distance between interest labels becomes smaller and potentially subtler, leading to increased proportions of disagreements and ties.

\subsection{Agreement with Recommender System Rankings}

We next examine how rankings of different recommender systems compare when derived from  human labels versus the LLM-judge predictions. To assess this, we compute the Kendall’s $\tau$ between the ranking of 14 recommender systems using the human relevance labels and the corresponding rankings obtained from LLM-judge–predicted labels, considering the full scale of interest levels from 1--7 to calculate the Compatibility measure as recommended by \citet{smucker2025extending}. We also include a weighted variant that emphasizes disagreements involving top-ranked items.

As shown in Table~\ref{tab:kendall_tau_pivoted}, increasing the number of LLM-judge predictions per user (selecting items to judge randomly) from 10 to 100 substantially improves the ranking agreement with the complete set of human relevance labels. Beyond 100 predictions per user, the improvement plateaus, indicating a performance ceiling, as extending predictions to 200 movies does not further increase agreement. With 100 LLM-judge predictions per user (corresponding to approximately 5,100 relevance labels---about \textasciitilde16\% of the total human-labeled set), we obtain a high Kendall’s $\tau$ of 0.92 for nDCG@100 between LLM-judge and the human judgments across 52 system configurations. This level of agreement is comparable to the values reported by~\citet{arabzadeh2025benchmarking} for ad hoc retrieval collections such as TREC-DL, which range between 0.77 and 0.92. These findings suggest that LLM-judge can approximate human-based model rankings with high fidelity, even when using a fraction of the human-labeled data.

\begin{table}[t]
\centering
\caption{Agreement between LLM-judge and human system rankings by the number of LLM-judge predictions per user.}
\renewcommand{\arraystretch}{1.15}
\setlength{\tabcolsep}{3pt}
\small
\begin{tabular}{c ccc ccc}
\toprule
 & \multicolumn{3}{c}{\textbf{Kendall's $\tau$}} & \multicolumn{3}{c}{\textbf{W-Kendall's $\tau$}} \\
\cmidrule(lr){2-4} \cmidrule(lr){5-7}
\textbf{K} & \textbf{Compat} & \textbf{nDCG} & \textbf{Bpref} & \textbf{Compat} & \textbf{nDCG} & \textbf{Bpref} \\
\midrule
10  & .710 & .831 & \textbf{.873} & .738 & .811 & .848 \\
50  & .802 & .899 & .855 & \textbf{.797} & .910 & .860 \\
100 & .802 & \textbf{.923} & .866 & .737 & \textbf{.913} & \textbf{.863} \\
200 & \textbf{.829} & .896 & .858 & .774 & .894 & .854 \\
\bottomrule
\end{tabular}
\label{tab:kendall_tau_pivoted}
\end{table}

\begin{figure}[t]
    \centering
    \includegraphics[width=1\linewidth]{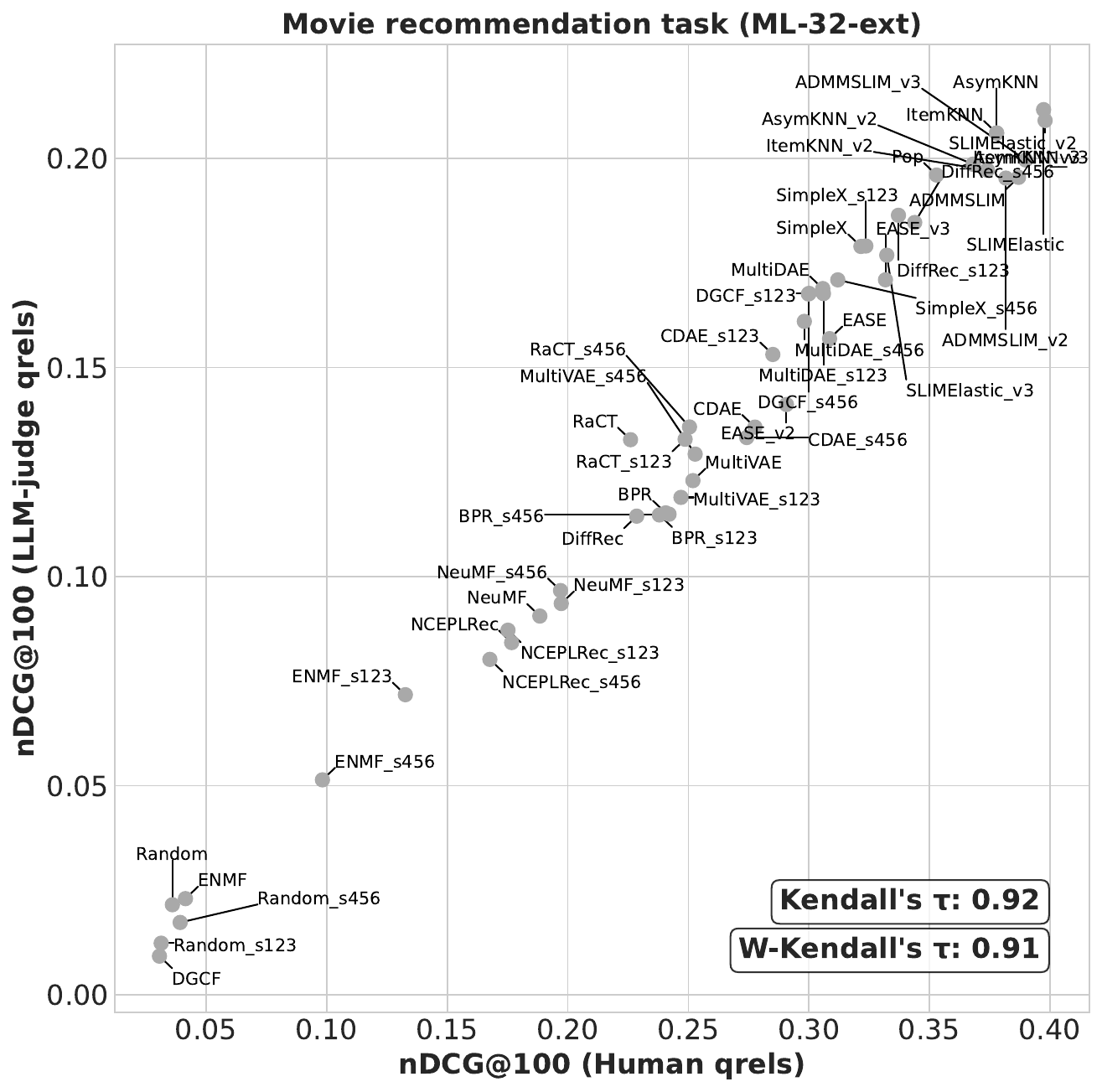}
    \caption{nDCG@100 correlation between LLM-judge (100 predictions per user) and human relevance labels for 52 system configurations.}
    \label{fig:ranking_cor}
\end{figure}

The per-system values shown in Figure~\ref{fig:ranking_cor} illustrate this relationship and reveal that the LLM-judge recovers system rankings that are distorted under traditional train-test evaluation (Section~\ref{sec:limitations}). The runs identified earlier as undervalued by the train-test split---SimpleX, which drops from rank 47 to 14 when moving to the Cranfield collection, and DGCF, which improves by up to 29 positions---are ranked at positions 16 and 23 respectively by the LLM-judge, closely matching their Cranfield ranks. Similarly, the overvalued NCEPLRec, ranked 1\textsuperscript{st} by train-test but 43\textsuperscript{rd} by Cranfield, is correctly placed at rank 44 by the LLM-judge. Across all 52 configurations, the median absolute rank displacement between LLM-judge and Cranfield is just 1 position, compared to 8 for the train-test split. This suggests that LLM-judge can identify undervalued systems that traditional evaluation methodologies systematically penalize due to label incompleteness.

\paragraph{\rev{Why moderate item-level agreement still yields stable system rankings}}
\rev{The contrast between moderate item-level alignment (Section~5.1) and high system-level ranking agreement warrants explanation. System-level metrics aggregate over thousands of user--item judgments, so item-level errors that are not systematically correlated with particular systems largely cancel out: what matters is not that every label is correct, but that errors do not consistently favor some systems over others---mirroring IR meta-evaluation, where assessor disagreement affects absolute scores but leaves system orderings largely intact~\cite{voorhees1998variations}. Two caveats follow. First, aggregation only protects against \emph{unbiased} noise: LLM errors that correlate with system characteristics (e.g., overrating popular items) can distort rankings in ways a global $\tau$ does not reveal. Second, stable rankings across 52 configurations do not guarantee reliable discrimination between closely matched top systems~\cite{otero2025limitations}. LLM-judge scores should thus be read as a system-level signal, not a substitute for item-level human ground truth.}

\paragraph{Sensitivity of rankings to unjudged items}

The experiments above restrict the LLM-judge to predicting labels only for items that were already judged by humans, treating all unjudged items as irrelevant. While this enables a controlled comparison with the human ground truth, it leaves open the question of how sensitive system rankings are to the large number of unjudged items in each system's retrieved list. To investigate this, we allow the LLM-judge to evaluate unjudged items from the top-100 retrieved results of each recommender system and compare system rankings under three evaluation scenarios:
\begin{itemize}
    \item \textbf{Human-only:} The current setup---only human-judged items are evaluated, unjudged items are treated as irrelevant.
    \item \textbf{LLM fills gaps:} Human labels are used where available; for unjudged items in the top-100, LLM-judge predictions are used instead.
    \item \textbf{Full LLM:} The LLM-judge evaluates all top-100 items, ignoring human labels entirely.
\end{itemize}

We note that without additional human judgments on the unjudged items, we cannot verify whether the LLM labels are correct. Instead, this experiment measures the \emph{stability} of system rankings when moving from sparse human judgments to denser LLM-augmented evaluations.

\begin{table}[t]
\centering
\caption{System ranking agreement (Kendall's $\tau$ on Compat) across evaluation scenarios and average Judged@100 for each. \textit{Human-only} treats unjudged items as irrelevant; \textit{LLM fills gaps} augments human qrels with calibrated LLM predictions for unjudged items; \textit{Full LLM} uses only LLM predictions.}
\renewcommand{\arraystretch}{1.15}
\setlength{\tabcolsep}{4pt}
\begin{tabular}{@{}lccc@{}}
\toprule
 & \textbf{Human-only} & \textbf{LLM fills gaps} & \textbf{Full LLM} \\
\midrule
vs.\ Human-only   & ---  & .844 & .763 \\
vs.\ LLM fills gaps & .844 & ---  & .797 \\
vs.\ Full LLM     & .763 & .797 & ---  \\
\midrule
Avg.\ Judged@100   & .559 & 1.00 & 1.00 \\
\bottomrule
\end{tabular}
\label{tab:mnar_scenarios}
\end{table}

As shown in Table~\ref{tab:mnar_scenarios}, filling unjudged gaps with LLM predictions increases the completeness of relevance labels (Judged@100) from 0.56 to 1.00. System rankings remain stable under these denser evaluations, with Kendall's $\tau$ = 0.84 between \textit{Human-only} and \textit{LLM fills gaps} scenarios. Notably, Judged@100 varies widely across individual systems (0.07--0.87), yet the overall system ordering is preserved once gaps are filled with calibrated LLM predictions.  The \textit{Full LLM} scenario, which replaces all human labels with LLM predictions, also maintains strong agreement ($\tau$ = 0.76 with \textit{Human-only}), further supporting the viability of LLM-only evaluation when human judgments are unavailable. For the \textit{LLM fills gaps} scenario, the LLM and human labels must be placed on a common scale before mixing, because the LLM assigns systematically higher scores than humans for the same items (mean 5.3 vs.~1.4 on the 0--7 scale). We calibrate LLM scores using the mean human score per LLM grade, estimated from the 13{,}336 items that have both labels. We report Compat in Table~\ref{tab:mnar_scenarios} rather than nDCG@100 because nDCG uses exponential gains ($2^{rel}-1$), so even small residual score differences between the two sources translate into large gain differences---for example, a relevance of 5 contributes 31$\times$ the gain of a relevance of 1. This makes nDCG rankings in the mixed-source \textit{LLM fills gaps} scenario unstable ($\tau$ = 0.26 with \textit{Human-only}). Compat, which compares systems by the fraction of item pairs ranked consistently with the relevance labels, is less sensitive to absolute score magnitudes and therefore more appropriate when combining heterogeneous label sources.

\rev{We emphasize the epistemic status of these results: the findings on human-judged items (Tables~\ref{tab:metadata_ablation} and~\ref{tab:kendall_tau_pivoted}) are \emph{directly validated} against human ground truth, whereas Table~\ref{tab:mnar_scenarios} constitutes \emph{indirect} evidence---rankings do not change erratically when gaps are filled, but the filled labels themselves are unverified. Validating them requires new human judgments, which we leave to future work.}

\section{Industrial Case Study: Applying LLM-judge for Model Selection}

We now examine LLM-judge in an industrial podcast recommendation setting, where evaluations based on logged interactions are biased toward the production model and Cranfield-style collections are impractical. We employ a criteria-based LLM-judge that predicts interest alignment between a user and a recommended podcast across multiple dimensions (topic, host \& guest, style \& format, and tone \& genre). Using 100 users, we score the top five recommendations per candidate model. Over a four-month development period, LLM-judge scores were computed alongside traditional offline metrics and human annotations to select the final model for A/B testing.

\paragraph{LLM-judge vs.~historical interaction metrics}
We compare LLM-judge rankings with traditional offline metrics derived from logged interactions (global time train-test splits).

\begin{figure}[t]
    \centering
    \includegraphics[width=1\linewidth]{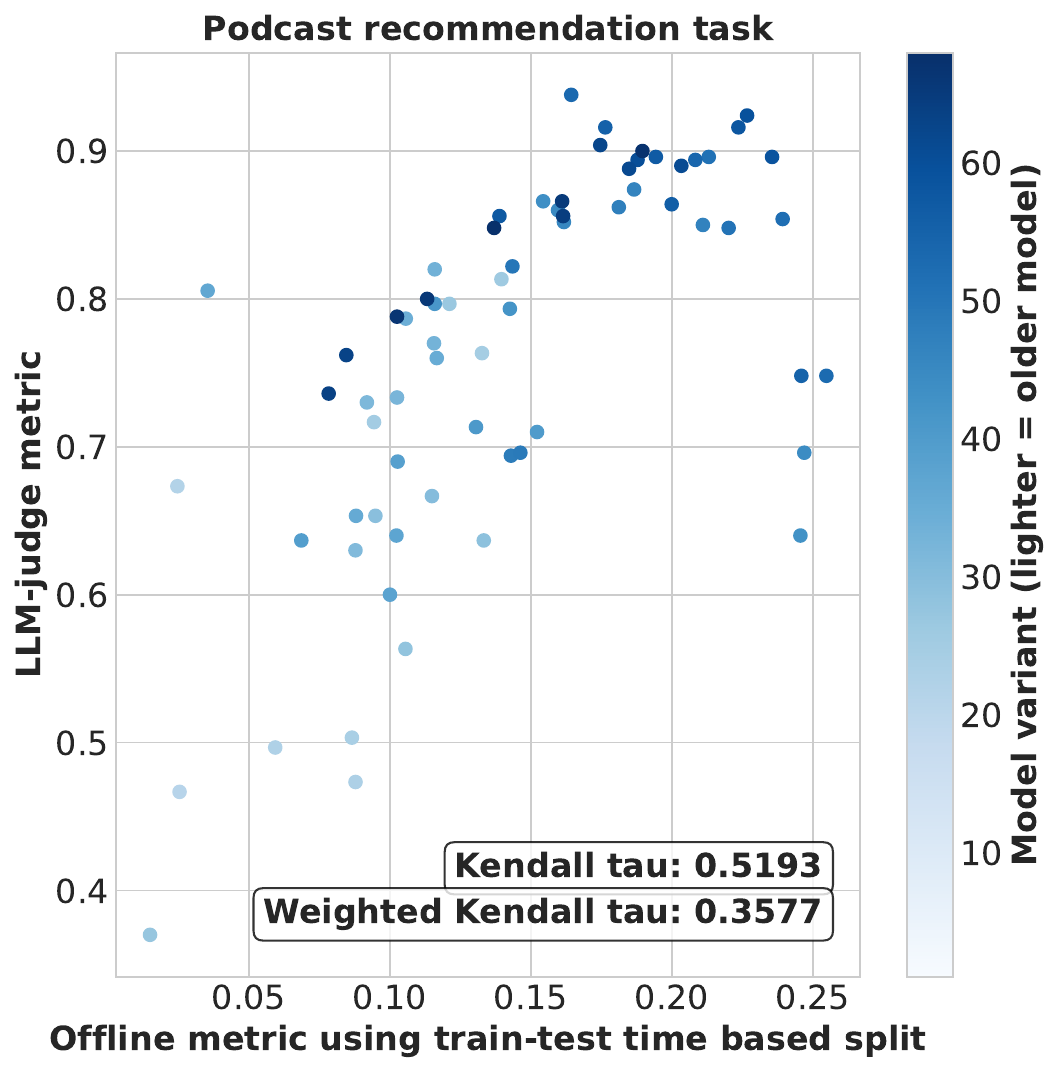}
    \caption{Correlation of recommender system rankings between LLM-judge and offline metrics based on historical interactions global train-test split.}
    \label{fig:ranking_cor_genrecs}
\end{figure}

As shown in Figure~\ref{fig:ranking_cor_genrecs}, the agreement between LLM-judge and conventional offline metrics is low. One explanation for this disparity is that LLM-judge mitigates exposure and popularity biases that arise from the incompleteness of relevance labels in historical interaction data. In fact, when evaluating a model that applied a popularity debiasing technique, we observed diverging behaviors between the two evaluation methods. Offline metrics based on logged interactions decreased notably, whereas LLM-judge scores increased. Consequently, the model performing best according to LLM-judge was not the same as the one ranked highest by traditional offline metrics.

\begin{table}[t]
\centering
\caption{Comparison between two podcast recommender system models using different evaluation metrics in the context of model selection.}
\renewcommand{\arraystretch}{1.15}
\setlength{\tabcolsep}{8pt}
\begin{tabular}{lcc}
\toprule
 & \textbf{Model 1} & \textbf{Model 2} \\
\midrule
Train--test time based split R@30 & \textbf{0.512} & 0.295 \\
LLM--judge avg. interest alignment & 0.734 & \textbf{0.890} \\
Employee preference & 0.353 & \textbf{0.647} \\
\bottomrule
\end{tabular}
\label{tab:model_comparison}
\end{table}

To further investigate, we conducted an internal evaluation with 20 employees, who compared the recommendations produced by two model variants. The results, summarized in Table~\ref{tab:model_comparison}, show that the model preferred by employees was the same as the one favored by LLM-judge (the variant incorporating a popularity debiasing mechanism), rather than the model ranked higher by conventional offline evaluation. These findings suggest that the LLM-judge can provide a less popularity biased assessment of model quality compared to historical interaction-based evaluations.\footnote{We acknowledge that, while this study illustrates the potential of LLM-judges, the number of participants is low and further research is necessary to investigate popularity debiasing effects in LLM-judges for recommendation}


\paragraph{Summary}
LLM-judge provided valuable complementary insights: while train--test metrics reflect how well models fit logged data, LLM-judge can evaluate recommendations for new objectives such as interest diversity. The selected model was deployed as an additional retrieval source in a large-scale streaming platform, increasing non-habit podcast episode streams\footnote{We define non-habit podcast streams as episode streams from shows for which the user did not listen to in the past 28 days.} by 3.5\% while keeping guardrail metrics intact.

\section{Discussion and Conclusion}\label{sec:discussion}

\paragraph{\rev{Risks of delegating assessment to LLMs}}
\rev{Delegating relevance assessment to LLMs carries risks that complete-looking label coverage can mask. First, LLM-judges can hallucinate: generated reasoning may confidently reference inaccurate plot elements, cast members, or user preferences, producing plausible but wrong labels~\cite{ni2025knowledge}---errors that, unlike missing human judgments, are silent and hard to audit at scale. Second, LLM-judges may inherit training-data biases: LLMs perform better and are better calibrated on popular knowledge~\cite{ni2025knowledge}, so judgments for long-tail or recent items---precisely where human labels are scarcest---are likely least reliable, though evidence on the direction of popularity effects in LLM-based recommendation is mixed~\cite{lichtenberg2024large}. Our case study offers encouraging but limited evidence here (the LLM-judge favored a popularity-debiased model, agreeing with human preferences); a systematic audit of long-tail effects remains open. Third, circularity arises when the evaluated systems themselves rely on LLMs~\cite{li2025preference}, and LLM-judges can be adversarially steered toward positive labels~\cite{10.1145/3673791.3698431}. Finally, alignment with human labels and ranking agreement on human-judged items are directly validated (Sections~5.1 and~5.2), whereas the LLM-judge's behavior on unjudged items, its resilience to the biases above, and its reliability in other domains are inferred indirectly. We therefore recommend LLM-judges for pre-screening, model triage, and augmenting sparse human labels, with human judgments retained as the ultimate arbiter.}

\paragraph{Judging vs.\ recommending} One might ask: if an LLM can judge relevance well enough, could it simply replace the recommender system itself? Judging relevance is a post-hoc evaluation of a single user--item pair given rich contextual information. Recommendation, by contrast, requires efficiently searching over large sets of candidate items under latency constraints, balancing exploration and exploitation, and optimizing for long-term engagement.

In this paper, we investigated the potential of LLMs to serve as automatic judges for recommender system evaluation, addressing long-standing challenges of scalability and incompleteness in existing offline methodologies. Our central question---whether the LLM-judge paradigm, validated in ad hoc retrieval, transfers to the subjective and profile-driven nature of recommendation---is directly relevant to the broader agenda of unifying search and recommendation evaluation. Using the \mlext{} Cranfield-style collection, we demonstrated that traditional historical train–test splits suffer from label sparsity, leading to unstable and unreliable model rankings. In contrast, the LLM-judge approach produces complete relevance estimations across systems, \rev{which can help mitigate} selection and exposure biases at a moderate cost\rev{, while introducing risks of its own, as discussed above}.

Our results show that the alignment between LLM-judge predictions and human labels improves when richer item metadata and longer user histories are provided\rev{, although item-level agreement remains moderate, particularly for fine-grained interest distinctions}. We further demonstrate that rankings derived from LLM-judge relevance labels achieve high agreement with human judgments (Kendall’s $\tau$ up to 0.92 for nDCG@100), comparable to those observed in ad hoc retrieval benchmarks\rev{---a contrast we attribute to system-level aggregation canceling out item-level errors that are not correlated with particular systems (Section~5.2)}. Crucially, the LLM-judge correctly recovers system rankings that are distorted under traditional evaluation, identifying systems that are systematically undervalued or overvalued due to label incompleteness. In addition, our industrial case study in the podcast domain illustrates that LLM-judge can assist model selection and evaluation in real-world settings, offering a scalable complement to human annotation and traditional offline metrics.

Looking ahead, several directions warrant exploration. First, incorporating human-in-the-loop frameworks could strengthen reliability and accountability, particularly given the risks of relying solely on automated judgments. Second, systematically studying biases introduced by LLM-judges, for example, through popularity, domain representation, or prompt design, will be critical for ensuring fairness and validity in evaluation. Third, prompt optimization---including explicit scale definitions and richer user profile representations---could further improve alignment; an ablation with explicit scale definitions showed that system-level rankings are robust to this choice (Kendall's $\tau$ differences $< 0.002$\rev{; Appendix~\ref{app:prompt_ablation}}). Finally, advancing human–LLM alignment through techniques such as in-context learning, supervised fine-tuning, or reinforcement learning could further enhance agreement and interpretability.

Overall, our findings suggest that the LLM-judge evaluation paradigm \rev{can transfer} from IR to RecSys \rev{as a promising \emph{complementary} evaluation signal---rather than a proven replacement for human or interaction-based evaluation---}representing a promising step toward scalable and robust evaluation frameworks for recommender systems, bridging the IR and RecSys evaluation traditions.







\appendix

\section{\texorpdfstring{\rev{Recommender System Model Hyperparameters}}{Recommender System Model Hyperparameters}}
\label{app:hyperparams}

\rev{Table~\ref{tab:hyperparams} lists the hyperparameters used for each recommender system model. Models marked with $\dagger$ participated in the original \mlext{} pooling phase; their hyperparameters follow~\citet{smucker2025extending}. All remaining models use RecBole default settings and are trained for 10 epochs. Full retrieval over the entire item catalog is performed for all models.}

\rev{To increase the number of systems for computing ranking agreement, we create two additional variants per model. For stochastic models (those involving neural network training), we vary the random seed (\texttt{seed} $\in \{123, 456\}$, base: 42), which changes weight initialization while keeping the data split fixed. For deterministic models (closed-form or neighbor-based), we vary a key hyperparameter: \texttt{reg\_weight} for EASE (250, 1000), $\lambda_1$/$\lambda_2$ for ADMMSLIM (2/500, 10/2000), $\alpha$/$\ell_1$ ratio for SLIMElastic (0.1/0.05, 0.5/0.1), and $k$ for ItemKNN and AsymKNN (30, 200).}

\begin{table}[H]
\centering
\caption{\rev{Model hyperparameters ($\dagger$ = pooled model).}}
\label{tab:hyperparams}
\small
\rev{\begin{tabular}{llrl}
\toprule
\textbf{Model} & \textbf{Type} & \textbf{Ep.} & \textbf{Key Hyperparameters} \\
\midrule
Pop$^\dagger$       & Non-pers.    & 1   & --- \\
Random              & Non-pers.    & 1   & --- \\
\midrule
EASE$^\dagger$      & Linear       & 1   & reg\_weight\,=\,500 \\
ADMMSLIM$^\dagger$  & Linear       & 1   & $\lambda_1$\,=\,5, $\lambda_2$\,=\,1000 \\
SLIMElastic         & Linear       & 1   & $\alpha$\,=\,0.2, $\ell_1$\,=\,0.02 \\
NCEPLRec            & Linear       & 10  & rank\,=\,450, $\beta$\,=\,1.0 \\
\midrule
BPR$^\dagger$       & MF           & 659 & emb\,=\,2048, lr\,=\,1e-4 \\
NeuMF$^\dagger$     & MF           & 137 & mlp\,=\,[128,64], lr\,=\,5e-4 \\
ENMF                & MF           & 10  & emb\,=\,64, neg\_w\,=\,0.5 \\
SimpleX             & MF           & 10  & emb\,=\,64, margin\,=\,0.5 \\
\midrule
CDAE$^\dagger$      & AE           & 176 & lr\,=\,0.05, reg\,=\,0.01 \\
MultiVAE$^\dagger$  & AE           & 227 & mlp\,=\,[300], lr\,=\,0.01 \\
MultiDAE            & AE           & 10  & mlp\,=\,[600], dropout\,=\,0.5 \\
RaCT                & AE           & 10  & mlp\,=\,[600], lat\,=\,256 \\
DiffRec             & AE           & 10  & steps\,=\,5, emb\,=\,10 \\
\midrule
ItemKNN$^\dagger$   & Neighbor     & --- & $k$\,=\,100 \\
AsymKNN             & Neighbor     & --- & $k$\,=\,100 \\
\midrule
DGCF                & Graph        & 10  & emb\,=\,64, factors\,=\,4 \\
\bottomrule
\end{tabular}}
\end{table}

\section{\texorpdfstring{\rev{Prompt Sensitivity: Explicit Scale Definitions}}{Prompt Sensitivity: Explicit Scale Definitions}}
\label{app:prompt_ablation}

\rev{As shown in Figure~\ref{fig:llm_prompt}, the baseline LLM-judge prompt instructs the model to predict interest on a scale of 0 to 7, but does not define what each level means. In contrast, the human annotators of \mlext{} received explicit definitions for each interest level during the labeling process~\cite{smucker2025extending}. To assess the sensitivity of the LLM-judge to this prompt design choice, we create a \textit{defined-scale} prompt variant that augments the baseline prompt with the following scale definitions, matching the instructions given to human annotators:}

\begin{quote}
\footnotesize
\rev{\texttt{0 = Not interested} \\
\texttt{1 = Somewhat interested} \\
\texttt{2 = Interested} \\
\texttt{3 = Very interested} \\
\texttt{4 = Extremely interested} \\
\texttt{5 = Ranked 3 (would place in their top-3 picks)} \\
\texttt{6 = Ranked 2 (would place in their top-3 picks)} \\
\texttt{7 = Ranked 1 (would place in their top-3 picks)}}
\end{quote}

\rev{All other aspects of the experimental setup remain identical: the same model (\textit{gpt-5-2025-08-07}), user profiles (1000 items), and metadata fields.}

\paragraph{\rev{Alignment with human labels (Section~5.1)}}
\rev{Table~\ref{tab:prompt_ablation_agreement} reports the pairwise agreement, tie, and disagreement rates for the baseline and defined-scale prompts. Adding explicit scale definitions slightly reduces the disagreement rate, but increases ties and lowers overall agreement. This suggests that the definitions cause the LLM to distribute scores more conservatively, clustering predictions around fewer distinct values.}

\begin{table}[t]
\centering
\caption{\rev{Effect of explicit scale definitions on pairwise agreement between LLM-judge and human labels (51 users, 50 recs/user).}}
\label{tab:prompt_ablation_agreement}
\renewcommand{\arraystretch}{1.15}
\setlength{\tabcolsep}{5pt}
\rev{\begin{tabular}{lccc}
\toprule
\textbf{Prompt variant} & \textbf{Agreement} & \textbf{Tie} & \textbf{Disagreement} \\
\midrule
Baseline       & 51.83\% & 29.97\% & 18.20\% \\
Defined scale  & 48.08\% & 34.88\% & 17.04\% \\
\bottomrule
\end{tabular}}
\end{table}

\paragraph{\rev{Agreement with system rankings (Section~5.2)}}
\rev{Table~\ref{tab:prompt_ablation_tau} reports Kendall's $\tau$ between the system rankings derived from LLM-judge labels and those derived from human labels, using 100 predictions per user across 52 system configurations. Both prompt variants yield nearly identical ranking agreement, with tau values differing by less than 0.002 for nDCG@100 and Compat. This demonstrates that the system-level ranking produced by LLM-judge is robust to the inclusion of explicit scale definitions in the prompt.}

\begin{table}[t]
\centering
\caption{\rev{Effect of explicit scale definitions on system ranking agreement (Kendall's $\tau$) with 100 predictions per user.}}
\label{tab:prompt_ablation_tau}
\renewcommand{\arraystretch}{1.15}
\setlength{\tabcolsep}{8pt}
\rev{\begin{tabular}{lccc}
\toprule
\textbf{Prompt variant} & \textbf{Compat} & \textbf{nDCG@100} & \textbf{Bpref} \\
\midrule
Baseline       & 0.8020 & 0.9229 & 0.8655 \\
Defined scale  & 0.7974 & 0.9214 & 0.8791 \\
\bottomrule
\end{tabular}}
\end{table}

\end{document}